\documentclass[fleqn,twoside]{article}
\usepackage{espcrc2}

\usepackage{graphicx}

\usepackage[figuresright]{rotating}

\newcommand{\AmS}{{\protect\the\textfont2
  A\kern-.1667em\lower.5ex\hbox{M}\kern-.125emS}}

\title{The Liverpool Telescope: Rapid  follow-up observation of Targets of Opportunity 
with a 2 m robotic telescope}

\author{Andreja Gomboc\address[ARI]{Astrophysics Research Institute,
        Liverpool John Moores University, \\ 
        12 Quays House, Egerton Wharf, Birkenhead, CH41 1LD, United Kingdom}, 
        \address[UL]{Department of Physics, University in Ljubljana, \\
        Jadranska 19, 1000 Ljubljana, Slovenia}
        Michael F. Bode\addressmark[ARI],
        David Carter\addressmark[ARI],
        Carole G. Mundell\addressmark[ARI],
        Andrew M. Newsam\addressmark[ARI],
        Robert J. Smith\addressmark[ARI]
        and
        Iain A. Steele\addressmark[ARI]}
       
\begin{document}

\begin{abstract}
The Liverpool Telescope, situated 
at Roque de los Muchachos Observatory, La Palma, Canaries, is the first 2-m, 
fully instrumented robotic telescope. It recently began observations. 
Among Liverpool Telescope's primary scientific goals is to monitor variable objects on all 
timescales from seconds to years. An additional benefit of its robotic operation is rapid 
reaction to unpredictable phenomena and their systematic follow up, simultaneous or coordinated 
with other facilities.
The Target of Opportunity Programme of the Liverpool Telescope includes the prompt search for and 
observation of GRB and XRF counterparts. A special over-ride mode implemented for GRB/XRF 
follow-up enables observations commencing less than a minute after the alert, including optical 
and near infrared imaging and spectroscopy. In particular, the moderate aperture and rapid 
automated response make the Liverpool Telescope excellently suited to help solving the mystery 
of optically dark GRBs and for the investigation of currently unstudied short bursts and XRFs.
\vspace{1pc}
\end{abstract}

\maketitle
\section{Introduction - robotic telescopes}

Robotic telescopes are, due to their efficiency and flexibility, opening
up numerous possibilities in many important branches of astrophysics \cite{Bode}.
Distinct advantages of robotic telescopes are:
\begin{itemize}
\item{
the rapid response to Targets of Opportunity (ToO) alerts,}
\item{the efficiency in the systematic monitoring of variable objects on all 
timescales from seconds to years,}
\item{ simultaneous and coordinated observations with
other facilities, either ground or space based,}
\item{conditions (seeing, photometricity, etc.) dependent observations, and}
\item{effective small scale surveys and routine tasks.}
\end{itemize}
They are invaluable in the study of variable astronomical objects of various types, 
including short and unpredictable phenomena, such as GRBs, XRFs, supernovae, dwarf 
novae and gravitational lensing; all types of binary systems and variable stars; 
and reaching from distant sources such as active galactic nuclei to the
study of comets and near-Earth asteroids in our Solar system.

With first light achieved at the end of July 2003, the Liverpool Telescope became
the world's largest operational robotic telescope and joins the international efforts 
in the study of exciting new fields of research in time domain astrophysics.
\begin{figure}
\centering
\includegraphics[height=5.2cm]{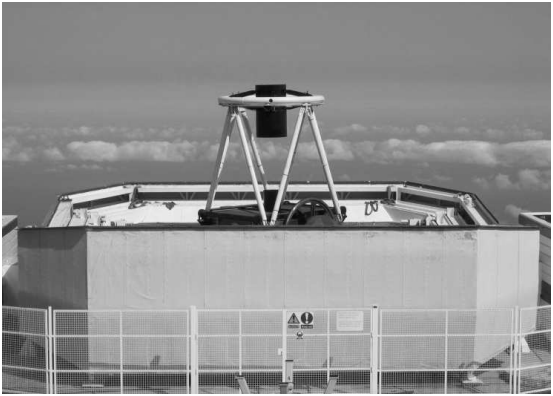}
\caption{The Liverpool Telescope at Roque de los Muchachos, La Palma, Canaries is 
a 2-m fully robotic altitude-azimuth design telescope with fully openable enclosure, 
which minimises the slewing time to the position of a ToO and enable observations 
starting less than a minute after the alert.}
\label{figLT}       
\end{figure}
\section{The Liverpool Telescope}
\subsection{Characteristics and instrumentation}
The Liverpool Telescope (LT) is the Altitude-Azimuth design telescope with a primary 
mirror diameter of 2 m and final focal ratio f/10 at the Cassegrain focus. Situated 
on La Palma it takes advantage of the excellent seeing on the site. Its enclosure is 
fully openable (Fig. \ref{figLT}), which minimises local thermal effects and the time 
to slew to new targets (with slewing rate of $>$2$^{\rm o}$/s). It can have five permanently 
mounted instruments (on 4 folded ports and one straight-through), which are selected 
by a deployable, rotating mirror. As listed in Table~\ref{tableLT}, the LT
instrumentation includes at present the RATCam optical CCD camera, with additions of 
the SupIRCam infra-red camera, low resolution Prototype spectrograph and FRODOSpec
Integral field double beam spectrograph in the near future. At present, the LT is 
still in the commissioning phase and for updated information please see 
http://telescope.livjm.ac.uk/.

\subsection{Robotic control and operating system}
As a fully robotic telescope with no night time supervision, the Liverpool Telescope
requires a Robotic Control System (RCS), which is designed  to act as a replacement 
for the duty astronomer. It is also required to be robust, reliable and adaptable to 
future instrument configurations and varying operational objectives \cite{FraserandSteele}. 
The RCS must include procedures such as telescope run-up, close down, focusing, 
scheduling and observing. In addition it must ensure that the telescope and any equipment 
is not damaged, therefore it must constantly monitor weather conditions and have the 
ability to respond to any fault and error in appropriate ways. It also needs to monitor 
the quality of the observations, i.e. the sky and seeing changes and conditions, since 
they influence the schedule of observing programme \cite{SteeleandCarter}. The LT possesses 
such a RCS.
\subsection{Automated scheduling}
Proposed observations are stored in a database and are selected from it with the scheduler,
which 
(i) must be sensible, so that observations which are impossible in current conditions 
should not be attempted,
(ii) efficient, so that observations are well matched to observing conditions and
the number of observations done is as high as possible, and
(iii) is fair to all partners according to the agreed percentage of telescope time.
 
For the LT, a relatively simple dispatch scheduler is chosen \cite{SteeleandCarter}, 
which does not attempt to make an optimum schedule for the whole night in advance
but merely looks for the best observation to do at a given time. The scheduler's 
decision is based on the current telescope state, Moon phase, observing conditions, 
object height, available time, scientific priority, urgency and fairness. 
\begin{table*}[htb]
\caption{The Liverpool Telescope instrumentation }
\label{tableLT}
\newcommand{\m}{\hphantom{$-$}}
\newcommand{\cc}[1]{\multicolumn{1}{c}{#1}}
\renewcommand{\tabcolsep}{2pc} 
\renewcommand{\arraystretch}{1.2} 
\begin{tabular}{@{}ll}
\hline
{\it RATCam} Optical CCD Camera - & 2048$\times$2048 pixels, 0.135"/pixel, FOV 4.6'$\times$4.6', \\
 & 8 filter selections (u', g', r', i', z', B, V, ND2.0) \\
 & - from LT first light, July 2003 \\
 \hline
 {\it SupIRCam} 1 - 2.5 micron Camera - & 256$\times$256 pixels, 0.4"/pixel, FOV 1.7'$\times$1.7', \\
  & Z, J, H, K' filters - from Autumn 2003 \\
  \hline
 {\it Prototype Spectrograph} - & 49, 1.7" fibres, 512$\times$512 pixels, R=1000; \\
  & 3500 $<$ $\lambda$ $<$ 7000 \AA - from Autumn 2003 \\
  \hline
 {\it FRODOSpec} Integral field  & R=4000, 8000; \\
  double beam spectrograph - & 4000 $<$ $\lambda$ $<$ 9500 \AA - from Summer 2004 \\
\hline
\end{tabular}\\[2pt]
\end{table*}
\section{ToO with the LT}
In general, a ToO proposal may be included in the database of the observations
on a timescale of 1 day, but for extremely time-critical objects such as for 
example GRBs, there is the possibility to use an Over-Ride mode built-in to the LT's
RCS. Upon receiving the ToO alert, the RCS automatically decides whether it is sensible 
to interrupt current observations and start the ToO follow-up. In the case of a positive 
outcome, it selects the appropriate instrument, begins automatic slew and starts 
observing the ToO region. Time elapsed between the alert and start of observation 
depends on the distance between the current telescope and ToO position, but should 
be on average within 30s since the receipt of the ToO alert. The instrumentation used 
and automated procedures can be suited to different characteristics and scientific 
points of interest regarding the type of the ToO.
\section{GRBs, XRFs and the LT}
High-priority ToO for the LT are GRBs and XRFs. 
Following the GRB/XRF alert from the GCN network and HETE-2, INTEGRAL and Swift (after its launch) 
we employ the Over-Ride mode and commence a search for and observation of GRB or XRF counterparts: 

(a) In the case of an alert with object error box larger than the FoV, we  
mosaic the entire error box (provided it is $<$30') with the Optical RATCam,
while in parallel, the automatic data processing and comparison with the 
catalogue (USNO-B1.0) is running to detect new objects in the field and to 
identify possible candidates for the GRB or XRF afterglow. 

(b) In alerts with error box smaller than the FoV we begin observations 
with optical multiband imaging of the field and search for the afterglow. 

After the successful commissioning of the SupIRCam and the spectrographs,
we will include near-infrared imaging in the first phases of observing, and, 
following the positive identification of the counterpart, continue observations 
with spectroscopy (provided that the afterglow magnitude is above the limiting 
magnitude, which is V$\sim$ 16 in 10 min for prototype and V$\sim$ 19 for FRODOSpec 
spectrographs respectively). With this routine as the starting point, automated 
procedures can be optimised with experience and adapted regarding scientific 
imperatives.  
\subsection{GRB and XRF science with the LT}
The discoveries of X-ray \cite{Costa} and optical transient \cite{Paradijs} sources 
associated with GRBs have revolutionised our understanding of the GRB phenomenon. 
While X-ray counterparts are now observed for essentially all GRBs, optical afterglows 
are detected in about half of them. Whether these missing OAs are inherently dark, dust 
absorbed, highly redshifted or just observationally overlooked is an open question \cite{Lazzati}.
So far, there are only 3 GRBs (GRB990123, GRB021004 and GRB021211) with optical afterglows 
detected within the first ten minutes after the GRB initial event. Their peak magnitudes 
lie between R$\sim$9 and $\sim$14 and they show rapid decay rates of 3-5 magnitudes in 10 min.
Given this rapid decline, it is easy to imagine that roughly 50\% of the bursts currently 
considered optically 'dark', may be detected in these early moments by a larger rapid response
telescope such as the LT. 

A vital clue to solve the dark burst problem probably lies in the infrared waveband, 
whether the bursts are dark due to high absorption by the dust in the close vicinity 
of a progenitor or further out in the host galaxy, or whether they are dark due to high 
redshift (z$>$10), which shifts the Ly$\alpha$ break to the infrared wavelengths.
Currently redshifts of about 40 GRBs are known \cite{jcg}, with the maximum of z=4.5 
detected for GRB 000131 \cite{Andersen}, but the selection is probably biased
towards bright and/or slowly decaying objects. In addition to GRBs redshift, 
spectra of GRB afterglows, especially in the early phases, hold a potential treasure of
valuable information about the environment and origin of the burst and will help 
distinguish among competing afterglow models for rebrightenings and color changes in 
the afterglows \cite{Lazzati}, 
\cite{Bersier}.

Of all the GRB afterglows so far observed, only one single epoch observation \cite{Castro-Tirado}
of a possible optical counterpart to a short GRB was reported. It has been predicted that 
the afterglows of short GRBs may be much fainter than the afterglows of long 
GRBs \cite{Panaitescu}. With an expected magnitude of less than 21 in the optical about 
1 hour after the burst a 2-m or even larger telescope is required for their rapid follow-up. 
Detection of the afterglows would be a break through in the study of short GRBs about which,
currently, very little is known. In the case of a non-detection, these observation will 
place more stringent limits on the afterglow magnitude immediately following the short GRB.

XRFs are recently identified class of phenomena with many characteristics of their prompt 
emission similar to long GRBs. They seem to be a natural extension to properties of GRBs 
and physical circumstances in which the explosion occurs \cite{Heise}. In the optical, 
several afterglows (XRF020903, XRF030723 \cite{Soderberg}, \cite{Fox}) and host galaxies 
\cite{Castro-Tirado2}, \cite{Fruchter} were detected. Detection of more optical afterglows 
and their early spectra will help unravel the origin of these events and their connection 
to GRBs.
\section{Conclusions}
The unique r\^ole that the robotic telescopes can play in the study of variable and transient 
sources has long been recognized. With the start of observations, the LT has great potential
for contributing to many interesting areas of time variable astrophysics. 
The short response time, moderate aperture, excellent site and LT's instrumentation
including optical camera, infrared imager and spectrograph, promise an interesting
and valuable scientific harvest from the ToO follow-up observations, especially afterglows 
of GRBs and XRFs. In collaboration with satelites (Swift, HETE-2, INTEGRAL) and
other ground-based facilities, including the Faulkes Telescopes (clones of the LT),
we expect to be able to follow-up around 1 in 6 GRB/XRF immediately after the alert 
and we plan to monitor scientifically interesting afterglows also into their later stages. 

\vspace{0.5cm}
{\it Acknowledgments} \\
The Liverpool Telescope is funded via EU, PPARC, JMU grants and the benefaction of Mr. A. E. Robarts.
A.G. acknowledges the receipt of the Marie Curie Fellowship from the EU.

\end{document}